\documentclass[twocolumn,twocolappendix]{aastex631}
\usepackage{bm}
\usepackage{amsmath}
\usepackage{amssymb}
\usepackage{graphicx}
\usepackage{float}
\usepackage{subfigure}
\usepackage{hyperref}
\usepackage{comment}
\usepackage{orcidlink}
\usepackage{ulem}
\usepackage{CJK}
\hypersetup{
	colorlinks=true,
	linkcolor=red,
	citecolor=blue,
}
\begin{document}

\title{Probing the equation of state of neutron stars with captured primordial black holes}

\author{Qing Gao\orcidlink{0000-0003-3797-4370}}
\email{gaoqing1024@swu.edu.cn}
\affiliation{School of Physical Science and Technology, Southwest University, Chongqing 400715, China}
\author{Ning Dai\orcidlink{0000-0002-0867-6764}}
\email{daining@hust.edu.cn}
\affiliation{School of Physics, Huazhong University of Science and Technology, Wuhan, Hubei
430074, China}
\author{Yungui Gong\orcidlink{0000-0001-5065-2259}}
\email{First author. gongyungui@nbu.edu.cn}
\affiliation{Institute of Fundamental Physics and Quantum Technology, Department of Physics, School of Physical Science and Technology, Ningbo University, Ningbo, Zhejiang 315211, China}
\affiliation{School of Physics, Huazhong University of Science and Technology, Wuhan, Hubei
430074, China}
\author{Chao Zhang\orcidlink{0000-0001-8829-1591}}
\email{Corresponding author. zhangchao1@nbu.edu.cn}
\affiliation{Institute of Fundamental Physics and Quantum Technology, Department of Physics, School of Physical Science and Technology, Ningbo University, Ningbo, Zhejiang 315211, China}
\author{Chunyu Zhang\orcidlink{0000-0002-4332-3729}}
\email{Corresponding author. chunyuzhang@yzu.edu.cn}
\affiliation{Center for Gravitation and Cosmology, College of Physical Science and Technology, Yangzhou University, Yangzhou 225009, China}
\author{Yang Zhao\orcidlink{0009-0003-7436-8668}}
\email{zhaoyangedu@hust.edu.cn}
\affiliation{School of Physics, Huazhong University of Science and Technology, Wuhan, Hubei
430074, China}

\begin{abstract}
Gravitational waves (GWs) from primordial black holes (PBHs) inspiraling within neutron stars (NSs)---should they exist---are detectable by ground-based detectors and offer a unique insight into the internal structure of NSs. To provide accurate templates for GW searches, we solve Einstein's equations within NSs and calculate the orbital motion of the captured PBH by considering dynamical friction, accretion, and gravitational radiation. Equipped with precise GW waveforms for PBHs inspiraling inside NSs, we find that the Einstein Telescope can differentiate between various equations of state for NSs. As PBHs inspiral deeper into NSs, the GW frequency rises near the surface, then decreases to a constant value deeper within NSs. The distinctive characteristics of GW frequency serve as the smoking gun for GW signals emitted by PBHs inspiraling inside NSs and can be used to probe the nuclear matter in the crust and core of NSs.
\end{abstract}



\section{Introduction}
Dark matter (DM) profoundly influences the structure and evolution of galaxies, clusters, and the cosmos as a whole, yet its exact nature and composition remain elusive \citep{Bertone:2004pz,Bertone:2018krk}.
Primordial black holes (PBHs) \citep{Zeldovich:1967lct,Hawking:1971ei,Carr:1974nx} generated by density fluctuations during inflation have emerged as a compelling candidate for DM \citep{Ivanov:1994pa,Khlopov:2004sc,Frampton:2010sw,Belotsky:2014kca,Clesse:2015wea,Carr:2016drx,Inomata:2017okj,Garcia-Bellido:2017fdg,Kovetz:2017rvv,  Ananda:2006af, Baumann:2007zm, Saito:2008jc,Alabidi:2012ex,Sasaki:2018dmp, Nakama:2016gzw, Kohri:2018awv, Cheng:2018yyr,Cai:2019amo, Cai:2018dig, Cai:2019elf, Cai:2019bmk,Cai:2020fnq, Pi:2020otn, Domenech:2020kqm, Liu:2018ess, Liu:2019rnx,Liu:2020cds,Liu:2021jnw,Liu:2022iuf,Yuan:2019fwv, Yuan:2019wwo,Yuan:2019udt,Papanikolaou:2021uhe,Papanikolaou:2022hkg,Chakraborty:2022mwu,Chen:2018czv,Chen:2018rzo,Chen:2019irf,Chen:2021nxo,Chen:2022fda, Zheng:2022wqo,Chen:2022qvg,Garcia-Bellido:2017mdw,Germani:2017bcs, Motohashi:2017kbs,Ezquiaga:2017fvi, Di:2017ndc, Ballesteros:2018wlw,Dalianis:2018frf, Kannike:2017bxn,Lin:2020goi,Lin:2021vwc,Gao:2020tsa,Gao:2021vxb,Yi:2020kmq,Yi:2020cut,Yi:2021lxc,Yi:2022anu,Kamenshchik:2018sig,Fu:2019ttf,Fu:2019vqc,Dalianis:2019vit,Gundhi:2020zvb,Cheong:2019vzl,Zhang:2021rqs,Kawai:2021edk,Cai:2021wzd,Chen:2021nio,Karam:2022nym,Ashoorioon:2019xqc}.
Observations of microlensing of stars in the Magellanic Clouds by nonluminous compact objects in the Galactic halo rule out PBHs in the mass range of
$10^{-7}\sim 10^2\ M_\odot$ making up most of the DM in the Galactic halo \citep{MACHO:1995lcj,MACHO:2000qbb,EROS-2:2006ryy,Paczynski:1985jf,Blaineau:2022nhy}.
However, it remains possible that a small fraction of the mass in a subpopulation of dark objects originates from PBHs,
and PBHs with masses in the range of $\left[10^{-16}, 10^{-10}\right]\ M_\odot$ may account for the total amount of DM \citep{Carr:2020gox}.
On the other hand, some gravitational wave (GW) events detected by Advanced Laser Interferometer Gravitational-Wave Observatory (LIGO) \citep{Harry:2010zz,LIGOScientific:2014pky}, Advanced Virgo \citep{VIRGO:2014yos} and Kamioka Gravitational Wave Detector (KAGRA) \citep{Somiya:2011np,Aso:2013eba} may support the existence PBHs \citep{Bird:2016dcv,Sasaki:2016jop,DeLuca:2021wjr,Franciolini:2021tla}.
Observational results from pulsar timing arrays (PTA) may also provide evidence for PBHs \citep{Vaskonen:2020lbd,DeLuca:2020agl}.
PBHs could even act as Planet 9, a hypothetical astrophysical object in the outer solar system that explains the anomalous orbits of trans-Neptunian objects \citep{Scholtz:2019csj}.
PBHs with subsolar mass in low mass ratio binaries can be constrained by the LIGO-Virgo-KAGRA data \citep{Phukon:2021cus},
and sub-solar-mass PBHs in extreme mass ratio inspirals (EMRIs) are potentially detectable \citep{Barsanti:2021ydd} by third-generation ground-based GW detectors, such as the Einstein Telescope (ET) \citep{Punturo:2010zz} and Cosmic Explorer (CE) \citep{Evans:2021gyd},
as well as space-based GW observatories like the Laser Interferometer Space Antenna (LISA) \citep{Audley:2017drz}, Taiji \citep{Hu:2017mde}, TianQin \citep{Luo:2015ght,Gong:2021gvw}, and Deci-hertz Interferometer GW Observatory (DECIGO) \citep{Kawamura:2011zz}.

It is possible for a sub-solar-mass PBH and a neutron star (NS) to form an EMRI system. NSs are compact objects with a mass of around 1--2$M_\odot$ and a radius of approximately 10--12 km.
In addition to the dominant radio pulsars,
there are other NS populations, including rotating radio transients, magnetars, X-ray dim isolated NSs, central compact objects,
and the more common X-ray NSs in binary systems that accrete  material from low-mass or high-mass companion stars \citep{Ascenzi:2024wws}.
From the gaseous atmosphere with a thickness of $\sim 10$ cm in the outermost NS layer,
to the ocean or envelope layer with a depth of $\sim$1--100 m composed of either heavy elements (e.g., iron) or light elements in a Coulomb liquid state,
to the solid crust layer with a thickness of $\sim 1$ km, and finally,
to the core with a radial size of around 10 km deeper inside the NS \citep{Ascenzi:2024wws}.
In the outer crust, the atoms are fully ionized.
At the neutron-drip density $\rho_\text{ND}\approx 4.3\times 10^{11}$ g cm$^{-3}$, neutrons begin to drip out from the nuclei, marking the start of the inner crust.
As the density increases to about $0.5 n_0$, a state of pure nuclear matter emerges, leading us to the NS core, where the nuclear saturation density  is $n_0=0.16$ nucleons per fm$^{3}$, or a mass density of $\rho_0\approx 2.8\times 10^{14}$ g cm$^{-3}$ \citep{Baym:2017whm}.
In the core regions,
quark matter may appear at high-density regimes in NSs.
The massive NSs with masses $\sim 2M_\odot$, such as PSR J1614-2230 \citep{Demorest:2010bx},
PSR J0740+6620 \citep{Fonseca:2021wxt} and PSR J0348+0432 \citep{Antoniadis:2013pzd}, have the inferred core density $\gtrsim$3--4$n_0$,
where a gradual onset of quark degrees of freedom and a transitional change from hadronic matter to quark matter are expected. Therefore, NSs serve as a cosmic laboratory for exploring the phases of cold dense strongly interacting nuclear matter.

The equation of state (EOS) of an NS is essential for understanding the internal structure of NSs, nuclear physics,
and the fundamental principles of quantum chromodynamics (QCD).
In principle, NS EOSs should be derived from the first principles of QCD.
The unified EOS is generated by a single effective nuclear Hamiltonian and is applicable throughout all regions within an NS.
Consequently, the transitions between the outer and inner crusts, as well as between the inner crust and the core, are treated consistently using the same physical model.
In \cite{Potekhin:2013qqa}, analytical representations of three unified EOSs for cold catalyzed nuclear matter developed by the Brussels--Montreal group: BSk19, BSk20, and BSk21 \citep{Goriely:2010bm,Pearson:2011zz,Pearson:2012hz} were proposed.
The model was then extended to include five new models: BSk22,  BSk23,  BSk24,  BSk25, and BSk26 \citep{Goriely:2013xba}.
Unfortunately, the BSK19 EOS does not support $2M_\odot$ NSs.
To account for the quark degrees of freedom in the core of NSs near their maximum mass,
the NS EOS QHC21
allowed for a transition region from nuclear to quark matter within the framework of quark--hadron crossover (QHC) \citep{Kojo:2021wax}.
The models BSk21 and BSk26 are part of the Brussels--Montreal series of nuclear energy density functionals, which predict similar overall structures for NSs but differ in detail.
The primary difference between BSk21 and BSk26 lies in crustal properties, such as the thickness of the crust.
In contrast, the main difference between the models BSk26 and QHC21AT lies inside the core of NSs.
QHC21AT is associated with a quark--hadron crossover EOS that features a more complex interior structure compared to the purely hadronic model BSk26 within the NS core.
Outside the core, there is little difference between the models BSk26 and QHC21AT.
For the EOS models BSk21, BSk26, and QHC21AT, the radii of the NS with the mass $M_\text{NS}=1.4\ M_\odot$ are 12.59, 11.77, and 11.84 km, respectively \citep{Potekhin:2013qqa,Goriely:2013xba,Kojo:2021wax}.
However, the EOS of NSs remains poorly understood due to a lack of direct observations.
Accurate determination of the mass and radius of NSs through observations like the Neutron Star Interior Composition Explorer can provide strong constraints on the NS EOS and the physics under extreme conditions.
Nevertheless, the methods used to measure the mass--radius relation are typically based on a range of assumptions, leading to significant uncertainties and degeneracies in the measured values.

The detection of the first binary NS merger, GW170817 \citep{LIGOScientific:2017vwq},
and its electromagnetic counterparts heralded the beginning of a new era in multimessenger astronomy,
offering a groundbreaking opportunity for the study of NSs. The tidal deformability of binary NSs is highly sensitive to their compactness \citep{Baym:2017whm},
and the tidal deformability estimates
derived from binary NS mergers can be utilized to constrain NS EOSs \citep{Flanagan:2007ix,Perot:2019gwl,Chatziioannou:2020pqz,Gupta:2022qgg,Puecher:2023twf}.
Observations from the first binary NS merger, GW170817, already gave constraints on the EOSs of NSs in the intermediate density region \citep{Annala:2017llu,LIGOScientific:2018cki}.
By simulating GWs from binary NS mergers using the QHC EOS, the possibility of testing hadron--quark transitions and QCD physics was explored \citep{Huang:2022mqp,Fujimoto:2022xhv,Prakash:2023afe,Harada:2023eyg}.

If PBHs with a mass less than $10^{18}$ g constitute the majority of DM,
EMRIs with PBHs orbiting around NSs are expected to form \citep{Rahvar:2023iup}.
For NSs born from massive stars near the Galactic center, essentially every such star will experience collisions with compact dark objects,
including PBHs with mass $m_D=10^{-8}~M_\odot$ or less,
during the star's main-sequence lifetime \citep{Horowitz:2019aim}.
A total rate of PBH--NS collsions was estimated to be $10^{-5}$$-$$10^{-4}$ yr$^{-1}$ in the Galactic center \citep{Capela:2013yf, Huang:2022mqp}, $\approx 10^{-5}(10^{-8}M_\odot/m_D)$ yr$^{-1}$ \citep{Horowitz:2019aim} or $\sim 10^{-8}$ yr$^{-1}$ \citep{Genolini:2020ejw} in the Galaxy.
The event rate of these collisions, GWs,
and the corresponding electromagnetic emissions
provide unique insights into the populations of PBHs,
the interior of NS, and the particle properties of DM \citep{Capela:2012jz,Capela:2013yf,Kouvaris:2013kra,Pani:2014rca,Fuller:2017uyd,Abramowicz:2017zbp,East:2019dxt,Genolini:2020ejw,Horowitz:2019aim,Zou:2022wtp,Rahvar:2023iup,Markin:2023fxx}.
The EOS of dense matter can potentially be probed by LIGO through the difference between GWs from a PBH inspiraling inside a nonrotating strange star described by the simple bag model with massless quarks \citep{Farhi:1984qu} and a PBH inspiraling inside a nonrotating NS described by the hadronic EOS BSk24 \citep{Pearson:2018tkr}.

The hypothetical and speculative scenario in which a PBH is captured by an NS is quite intriguing \citep{Horowitz:2019aim,Zou:2022wtp,Baumgarte:2024mei,Baumgarte:2024buu,Baumgarte:2024iby,Baumgarte:2024ouj,Chen:2024qke}.
During the rapid growth of the PBH, the PBH drag manifests as Bondi accretion in the subsonic regime \citep{Genolini:2020ejw}.
The Bondi accretion of the NS onto the PBH leads to the destruction of the NS in a short time,
meaning that observations of NSs can impose constraints on the abundance of PBHs in the mass range of $10^{20}$$-$$10^{23}$ g \citep{Capela:2012jz,Capela:2013yf}.
The effects of NS rotation and the viscosity of nuclear matter on the accretion and spin evolution of PBHs were examined in \cite{Kouvaris:2013kra}.
Using fully general relativistic simulations,
it was found that PBHs inside NSs can acquire a nonnegligible spin and induce differential rotation in the NS core \citep{East:2019dxt}.
Assuming a homogeneous NS structure and subsonic motion of PBHs inside NSs,
it was shown that both the frequency and amplitude of the emitted GWs are constant \citep{Genolini:2020ejw}.
The orbital frequency and lifetime of a DM object in a circular orbit of radius $r$ inside a compact star are approximately $\nu=841\sqrt{\rho_c/(10^{14}\ \text{g\ cm}^{-3})}$ Hz and $T=68\left(M_\odot/m_D\right)\left(1 \text{m}^2/r^2\right)\left(3.3\text{ kHz}/f\right)^4$ h \citep{Horowitz:2019aim}, respectively.
Consequently, GWs from the oscillation of PBHs with masses $m_D\gtrsim10^{-8}\ M_\odot$ inside NSs with central density $\rho_c\approx 10^{15}\ \text{g/cm}^{3}$ exhibit typical frequencies $f_\text{GW}=2\nu$ in the range of 3--5 kHz,
which are detectable by ground-based detectors \citep{Horowitz:2019aim}.
By considering the Newtonian motion of a PBH inside an NS,
accounting for the dynamical friction of the NS medium, the drag force due to accretion,
and the energy loss from gravitational radiation,
it was found that the inspiraling of PBHs with mass $\sim 10^{-5}M_\odot$ inside NSs at a distance of 10 kpc can be detected with Advanced LIGO \citep{Zou:2022wtp}.
For eccentric orbits, the maximum amplitude of GWs emitted by PBHs captured inside NSs oscillates between the two polarizations,
resulting in quasiperiodic GW beats due to orbital precession \citep{Baumgarte:2024mei,Baumgarte:2024buu,Baumgarte:2024iby,Baumgarte:2024ouj,Chen:2024qke}.

Since GWs from PBHs inspiraling within NSs encode information about the internal structure of NSs and are detectable by ground-based GW observatories,
the EMRIs comprising a sub-solar-mass PBH and an NS can be utilized to probe the NS EOS,
the internal NS structure, and the physics under extreme conditions.
To realize this concept, it is essential to model the GW waveforms for the EMRIs accurately.
Numerical-relativity simulations of a black hole with a mass of $0.5~M_\odot$ captured by an NS with a mass of $1.4~M_\odot$
indicate that current GW models fail to accurately characterize the system \citep{Markin:2023fxx}.
We aim to provide accurate GW waveforms for sub-solar-mass PBHs inspiraling inside NSs.
For the first time, we solve Einstein equations to obtain the metric inside the NS and then numerically calculate the orbital motions of PBHs inside the NS by taking the dynamical friction, accretion, and GW radiation into account.
We then calculate GWs from PBHs inspiraling inside an NS using three different EOS models: BSk21, BSk26 and QHC21AT,
to investigate the internal NS structure and the capability of ground-based detectors to distinguish between the EOS models for NSs.
Since we are concerned about a PBH inspiraling inside an NS after a long inspiral evolution outside the NS,
we focus on a circular orbit only.
We use the units $G=c=1$.

\section{The metric and orbital motion}
\begin{figure}
\includegraphics[width=0.9\columnwidth]{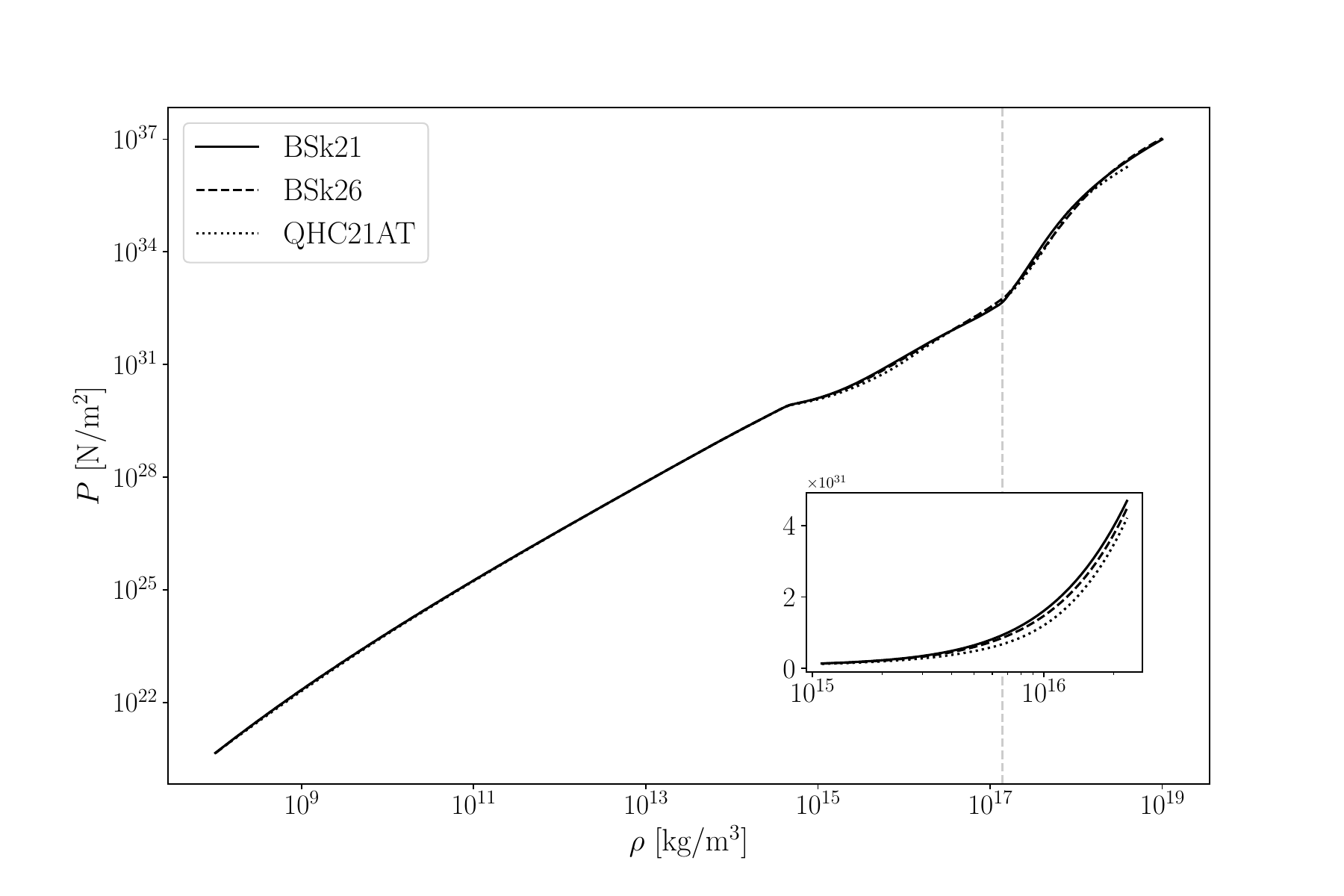}
\caption{
Pressure as a function of density for the EOS models BSk21 (solid line), BSk26 (dashed line), and QHC21AT (dotted line).
The vertical dashed line marks the density of $0.5 \rho_0$.
}
\label{Fig:p_rho}
\end{figure}

Assuming the NS is characterized by a perfect fluid and described by the static spherically symmetric metric
\begin{equation}
\label{eq:NS_metric}
ds^2=-e^{2\alpha(r)}dt^2+e^{2\beta(r)}dr^2+r^2(d\theta^2+\sin^2\theta d\phi^2),
\end{equation}
and introducing the mass function,
\begin{equation}
\label{massfun}
m(r)=\frac{1}{2}\left[r-re^{-2\beta(r)}\right]
\end{equation}
which represents the mass enclosed within radius $r$,
Einstein field equations give
\begin{equation}
\label{eq:NS_GR}
\begin{split}
	\alpha'=\frac{d\alpha}{dr}=&\frac{m(r)+4\pi r^3p}{r[r-2m(r)]},\\
	m'=\frac{dm}{dr}=&4\pi r^2\rho,
\end{split}
\end{equation}
where $\rho$ and $p$ are the density and pressure for perfect fluids.
The Tolman--Oppenheimer--Volkoff equation gives
\begin{equation}
\label{eq:NS_TOV}
-\frac{dp}{dr}=(\rho+p)\frac{m(r)+4\pi r^3p}{r[r-2m(r)]}.
\end{equation}
To solve Equations \eqref{eq:NS_GR} and \eqref{eq:NS_TOV},
we need to specify the NS EOS $P(\rho)$.
\cite{Potekhin:2013qqa} parameterizes the NS EOS as
\begin{equation}
\label{eq:NS_EoS}
\begin{split}
    \zeta=&\frac{a_1+a_2\xi+a_3\xi^3}{1+a_4\xi}\{\exp[a_5(\xi-a_6)]+1\}^{-1}\\
    &+\frac{a_7+a_8\xi}{\exp[a_9(a_6-\xi)]+1}\\
    &+\frac{a_{10}+a_{11}\xi}{\exp[a_{12}(a_{13}-\xi)]+1}\\
    &+\frac{a_{14}+a_{15}\xi}{\exp[a_{16}(a_{17}-\xi)]+1}\\
    &+\frac{a_{18}}{1+[a_{19}(\xi-a_{20})]^2}+\frac{a_{21}}{1+[a_{22}(\xi-a_{23})]^2},
\end{split}
\end{equation}
where $\xi=\log(\rho/\text{g\,cm}^{-3})$, $\zeta=\log(p/\text{dyn\,cm}^{-2})$,
and the coefficients $a_i$ for the EOS models BSk21 and BSk26 are given by \cite{Potekhin:2013qqa}, \cite{Goriely:2013xba}, and \cite{Pearson:2018tkr}.
The numerical result of the EOS model QHC21AT can be found in \cite{Kojo:2021wax}.
Figure \ref{Fig:p_rho} shows the pressure as a function of density for the EOS models BSk21, BSk26, and QHC21AT.
As shown in figure \ref{Fig:p_rho},
the pressure increases more rapidly with density when it reaches
$0.5\rho_0$,
with the QHC21AT model exhibiting the lowest pressure at the same density.

\begin{figure}
\includegraphics[width=0.99\columnwidth]{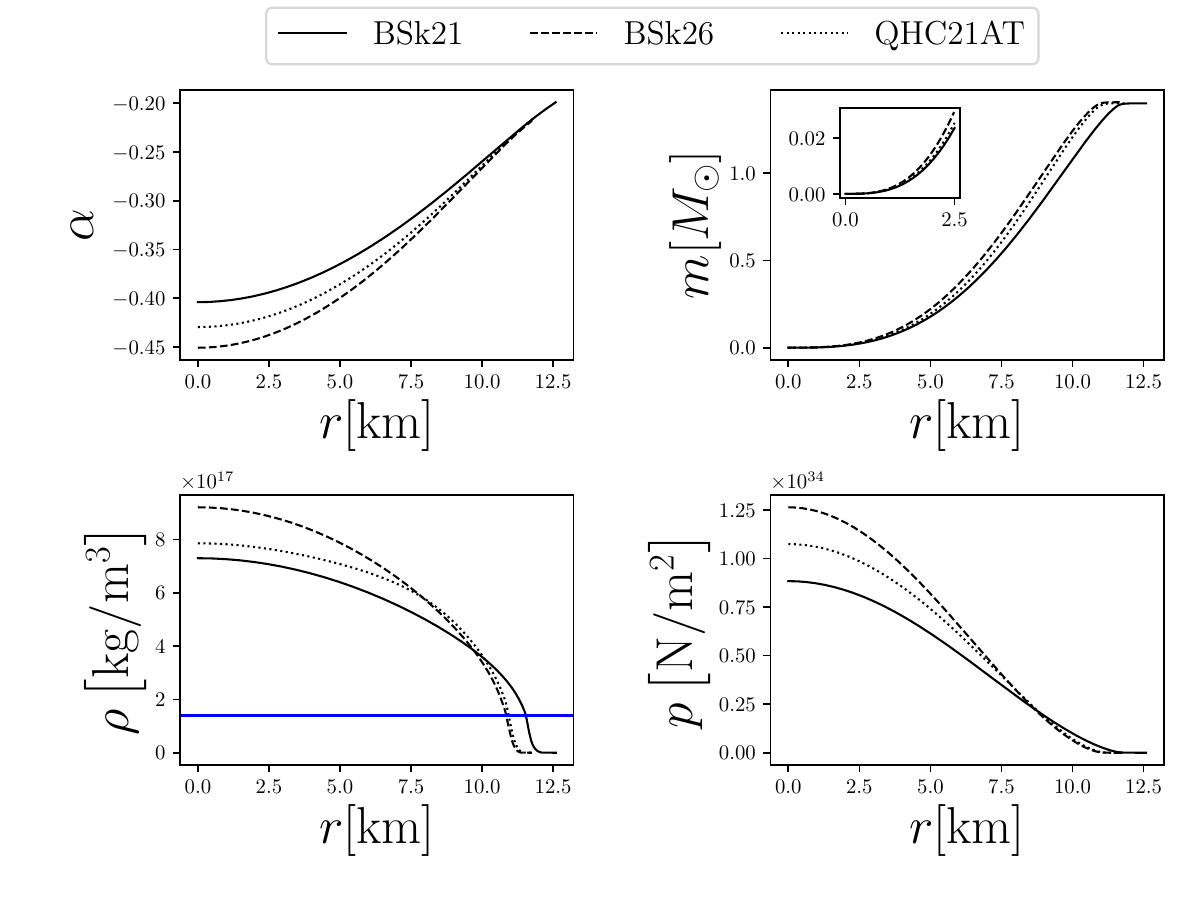}
\caption{
The metric parameter $\alpha(r)$, the mass function $m(r)$, and the density and pressure distributions for different EOS models, BSk21 (solid line), BSk26 (dashed line), and QHC21AT (dotted line).
The blue horizontal line in the lower left panel indicates the density of $0.5 n_0$.}
\label{Fig:NS_metric}	
\end{figure}

At the NS surface, we take the metric to be Schwarzschild metric and
set the energy density equal to that of $^{56}$Fe at zero pressure and zero temperature \citep{Potekhin:2013qqa}, i.e.  $\alpha(R)=\ln(1-2 M_\text{NS}/R)/2$, $\beta(R)=-\ln(1-2M_\text{NS}/R)/2$ and $\rho(R)=7.86\  \text{g\,cm}^{-3}$.
With these initial conditions,
we numerically solve the Einstein equation inside the NS.
For the first time, we obtain the space-time geometry inside the NS, as shown in Figure \ref{Fig:NS_metric}.

Outside an NS, the mass function $m(r)$ remains constant and is equal to the total mass of $M_\text{NS}=1.4 M_\odot$.
In the crust beneath the ocean, the atoms are fully ionized under intensive pressure;
the nuclei become increasingly neutron-rich as more electrons are captured,
and the density rises until it reaches $0.5\rho_0$.
As seen from Figure \ref{Fig:NS_metric},
near the NS surface,
the metric parameter $\alpha(r)$ and the mass functions $m(r)$ for three EOS models all approach approximately $-0.2$ and $1.4\ M_\odot$, respectively, which is consistent with the Schwarzschild metric at the surface.
The transitions from the inner crust to the core occur at $\{11.5,10.8,10.9\}$ km for the EOS models BSk21, BSk26, and QHC21AT,
respectively.
Since the density and pressure outside the NS core are much lower than those within,
the mass contributions from the NS crust and envelope ($r\gtrsim 10$ km) to the total mass are negligible,
and the mass outside the NS core remains constant as $r$ increases,
along with the orbital frequency $\Omega\sim\sqrt{Gm(r)/r^3}\propto r^{-3/2}$ outside the NS core.
In the NS core, there are primarily neutrons,
with a small fraction of protons and electrons (and at higher densities, also muons) to maintain charge neutrality.
As density and pressure increase with decreasing $r$,
both $\alpha(r)$ and $m(r)$ decrease.
Deep inside the NS core,
the density can reach up to $\sim 10^{18}\text{ kg/m}^3$.
At such extreme densities, understanding the material composition becomes even more uncertain,
and new particle species such as hyperons or even
transitions to more exotic states of matter like a Bose-Einstein
condensate of pions, and/or kaons, or a quark--gluon plasma, may emerge.
In the inner core ($r\lesssim3$ km),
the density, pressure, and $\alpha(r)$ are almost constant,
but the mass function $m(r)$ continues to decrease as $\sim r^3$.
In this region, $\alpha(r)$, $m(r)$, $\rho(r)$, and $p(r)$ exhibit the largest differences for different EOS.
Therefore, we expect to distinguish different EOSs through observations of GWs.
According to Newtonian gravity, if the density of the NS $\rho$ is constant,
then inside the NS the mass function $m(r)=4\pi\int \rho(r)r^2 dr$ equals $4\pi\rho r^3/3$,
and the orbital frequency $\Omega$ of a PBH in a circular orbit remains constant, as discussed in \cite{Genolini:2020ejw}.
In the Newtonian limit, $g_{tt}\approx -1+2Gm/r$,
but Figure \ref{Fig:NS_metric} shows that $|\alpha(r)|$ is not small and $g_{tt}=-e^{2\alpha}$ significantly differs from $-1$.
The fully relativistic results obtained from solving Einstein's equations diverge entirely from those in Newtonian gravity;
thus, the orbital motion and GW waveform will also differ significantly.
These results indicate that it is essential to solve the spacetime geometry within an NS to investigate the NS EOS,
as well as the internal NS structure and physics under extreme conditions.

\subsection{Energy loss}
When a DM object moves through a collisionless medium at speed $v$,
the gravitational pull from the wake of the DM object causes it to slow down; this effect is known as dynamical friction.
For a circular orbit, the dynamical friction force is \citep{Kim:2007zb,Zou:2022wtp}
\begin{equation}
\label{eq:F_DF}
 \bm{F}_{\rm DF}=-\frac{4\pi\rho m_D^2}{v^2}(I_r\hat{\bm{r}}+I_\phi\hat{\bm{\phi}}),
\end{equation}
where
\begin{equation}
\label{eq:Ir}
\begin{split}
 &I_r=\begin{cases}
		\mathcal{M}^2 10^{3.51\mathcal{M}-4.22},& \\
  \qquad\qquad\qquad\qquad\text{if }\mathcal{M}<1.1,& \\
		0.5\ln[9.33\mathcal{M}^2(\mathcal{M}^2-0.95)],&\\
   \qquad\qquad\qquad\qquad\text{if } 1.1\le \mathcal{M}<4.4,& \\
		0.3\mathcal{M}^2,& \\
   \qquad\qquad\qquad\qquad\text{if }\mathcal{M}\ge 4.4,&
 	\end{cases}
\end{split}
\end{equation}
\begin{equation}
\label{eq:Iphi}
\begin{split}
 &I_\phi=\begin{cases}
 		\mathcal{M}^3/3 - 0.80352 \mathcal{M}^4 + 7.68585 \mathcal{M}^5,&\\
    \qquad\qquad\qquad\quad~\text{if }\mathcal{M}<0.08588, &\\
	0.7706\ln\left(\frac{1+\mathcal{M}}{1.0004-0.9185\mathcal{M}}\right)-1.4703\mathcal{M},&\\
  \qquad\qquad\qquad\quad~\text{if } 0.08588<\mathcal{M}<1.0, & \\
	\ln[3300(\mathcal{M}-0.71)^{5.72}\mathcal{M}^{-9.58}],&\\
    \qquad\qquad\qquad\quad~\text{if } 1.0\le \mathcal{M}<4.4,& \\
		\ln\left(\frac{10}{0.11\mathcal{M}+1.65}\right),&\\
 \qquad\qquad\qquad\quad~\text{if } \mathcal{M}\ge 4.4, &
	\end{cases}
\end{split}
\end{equation}
the Mach number is $\mathcal{M}=v/c_s$, $v$ is the speed of the DM object, and $c_s=\sqrt{dp/d\rho}$ is the sound speed.

When PBHs move inside the NS, they will accrete the surrounding material.
The accretion rate can be described by the Bondi–Hoyle–Lyttleton accretion \citep{Edgar:2004mk},
\begin{equation}
\label{eq:mDdot}
\dot{m}_D=\frac{4\pi\lambda\,\rho\,m_D^2}{(c_s^2+v^2)^{3/2}},
\end{equation}
where the overdot indicates the differentiation with respect to $t$,
and we choose $\lambda=0.707$ \citep{Kouvaris:2013kra}.
The variation of orbital energy due to the mass increase is \citep{Blachier:2023ygh}
\begin{equation}
\label{eq:Pa}
P_a=\dot{m}_D \sqrt{e^{2\alpha(r)}\left[1-r
\alpha^\prime(r)\right] }\,,
\end{equation}
where the prime indicates the differentiation with respect to $r$.

Using the quadrupole--octupole formula,
the energy loss rate due to GW emissions is written in terms of the multipole moments of the source as
\begin{equation}\label{eq:RR_def}
\begin{split}
	\left\langle \frac{d E}{dt} \right\rangle_\text{GW}=&-\frac{1}{5}
	\Big\langle \dddot{\mathcal{I}}^{jk}\dddot{\mathcal{I}}^{jk}+\\
	&\frac{5}{189}\ddddot{\mathcal{M}}^{jkl}\ddddot{\mathcal{M}}^{jkl}  +\frac{16}{9}\dddot{\mathcal{J}}^{jk}\dddot{\mathcal{J}}^{jk} \Big\rangle,
\end{split}
\end{equation}
where the time on the right-hand side should take the retarded time,
the symmetric-traceless parts of mass quadrupole $\mathcal{I}^{jk}$, mass octupole $\mathcal{M}^{jkl}$, and flux quadrupole $\mathcal{J}^{jk}$ are
\begin{equation}
\label{eq:multipole}
\begin{split}
	\mathcal{I}^{jk}&=m_D \left(x^j x^k- \frac{1}{3}\delta^{jk}r^2\right) \left[1+\frac{1}{2}v^2+\right.\\
 &\left. \qquad \left(-\frac{1}{2}-\frac{3}{4}\delta-\frac{7}{32}\delta^2\right)U\right],\\
	\mathcal{M}^{jkl}&=m_D\left[ x^j x^k x^l -\frac{1}{5}\left(\delta^{jk}x^l r^2+\delta^{kl}x^j r^2+\delta^{jl}x^k r^2\right)\right],\\
	\mathcal{J}^{jk}&=\frac{1}{2}m_D\left( \epsilon^j_{mn}x^m v^n x^k+\epsilon^k_{mn}x^n v^m x^j \right),
\end{split}
\end{equation}
respectively,
$\epsilon^j_{mn}$ is the permutation symbol, $U=[1-e^{2\alpha(r)}]/2$ and $\delta=e^{2\alpha(r)}-e^{2\beta(r)}$.
Taking the time derivatives of these multipoles, we get
\begin{equation}
\label{eq:multipole2}
\begin{split}	\dddot{\mathcal{I}}^{jk}\dddot{\mathcal{I}}^{jk}&=32 r^4 m_{D}^2\Omega^6+72r^4 \Omega^4 \dot{m}_D^2\\
&\quad -2 r^4 m_{D}^2 \Omega^6 \left[-16r^2 \Omega^2+\left(16+24\delta^2 \right)U\right],\\
	\ddddot{\mathcal{M}}^{jkl}\ddddot{\mathcal{M}}^{jkl}&=\frac{8202}{5}r^6 m_{D}^2 \Omega^8+\frac{14592}{5}r^6 \Omega^6 \dot{m}_D^2,\\
	\dddot{\mathcal{J}}^{jk}\dddot{\mathcal{J}}^{jk}&=\frac{1}{2}r^6 m_{D}^2 \Omega^8+\frac{9}{2}r^6 \Omega^6 \dot{m}_D^2,
\end{split}
\end{equation}
where $\Omega$ is the orbital angular speed,
and only the first time derivative of $m_D$ is kept.
Substituting Equation \eqref{eq:multipole2} into Equation \eqref{eq:RR_def},
we get the energy flux of GW radiation,
\begin{equation}
\label{eq:F_RR}
\begin{split}
	\left\langle \frac{d E}{dt} \right\rangle_\text{GW}&=-\frac{32}{5} r^4 m_{D}^2 \Omega^6\\
	&\quad+\frac{2}{5} r^4 m_{D}^2 \Omega^6 \left[-16r^2 \Omega^2+\left(16+24\delta^2 \right)U\right]\\
	&\quad-\frac{8}{45}r^6 m_{D}^2 \Omega^8-\frac{2734}{315}r^6 m_{D}^2 \Omega^8 \\
	 &\quad-\left(\frac{72}{5}r^4 \Omega^4+\frac{4864}{315}r^6 \Omega^6+\frac{8}{5}r^6 \Omega^6 \right)\dot{m}_D^2,
\end{split}
\end{equation}

\subsection{Orbital Evolution}
With the adiabatic approximation, the PBH follows a geodesic path in the background metric (Equation \eqref{eq:NS_metric}) during each revolution.
The dynamical friction, accretion, and GW reaction can be considered external forces that modify the geodesic motion.
The external forces acting on the PBH can be divided into conservative and dissipative forces, $f^\mu=f^{\mu}_{\rm cons}+f^\mu_{\rm diss}$.
For the conservative component, only the radial part of the dynamical friction in Equation \eqref{eq:F_DF} is nonzero,
with $f^t_\text{cons}=f^\theta_\text{cons}=f^\phi_\text{cons}=0$, and
\begin{equation}
\label{frcons1}
f^r_\text{cons}=\frac{a^r}{e^{2\alpha(r)}[1-r\alpha '(r)]},
\end{equation}
where $a^r=-4\pi\rho m_D I_r\hat{\bm{r}}/v^2$ is the radial part of dynamical friction per unit mass.
In the presence of conservative external force $f^r_\text{cons}$, the circular motion of PBH becomes
\citep{Pound:2007th}
\begin{equation}\label{eq:geodesic}
\begin{split}
	&\frac{d^2t}{d\tau^2}+ 2\alpha'(r) \frac{dt}{d\tau}\frac{dr}{d\tau}=0, \\
	&\frac{d^2r}{d\tau^2}+\left(\frac{dt}{d\tau}\right)^2 e^{2 \alpha (r)-2 \beta (r)} \alpha '(r)-r e^{-2 \beta(r)}\left(\frac{d\theta}{d\tau}\right)^2\\
	&\quad+\beta '(r)\left(\frac{dr}{d\tau}\right)^2
	-\left(\frac{d\phi}{d\tau}\right)^2 r \sin ^2(\theta) e^{-2 \beta (r)}=f^r_{\rm cons},\\
	&\frac{d^2\theta}{d\tau^2}+ \frac{2}{r}\frac{dr}{d\tau}\frac{d\theta}{d\tau}-\sin(\theta)\cos(\theta) \left(\frac{d\phi}{d\tau} \right)^2=0, \\
	&\frac{d^2\phi}{d\tau^2}+ \cot(\theta)\frac{d\theta}{d\tau}\frac{d\phi}{d\tau} +\left(\cot(\theta)\frac{d\theta}{d\tau}+\frac{1}{r}\frac{dr}{d\tau}  \right)\frac{d\phi}{d\tau}\\
	&\quad+\frac{1}{r}\frac{dr}{d\tau}\frac{d\phi}{d\tau}=0.
\end{split}
\end{equation}

Together with the normalization condition $g_{\mu\nu}u^{\mu}u^{\nu}=-1$,
we can solve Equation \eqref{eq:geodesic} and get the four-velocity of the PBH in the case of circular orbits,
\begin{align}
        \label{velocity-4}
	&u^\mu=\frac{dx^{\mu}}{d\tau}=\left[u^t,0,0,u^\phi\right],\\
        \label{velocity-t}
	&u^t=\frac{dt}{d\tau}=\sqrt{\frac{1-re^{2\beta(r)}f^r_{\rm cons}}{e^{2\alpha(r)}\left[1-r \alpha^\prime(r)\right]}}\,,\\
        \label{velocity-phi}
	&u^\phi=\frac{d\phi}{d\tau}=\sqrt{\frac{e^{2\beta(r)f^r_{\rm cons}}-\alpha^\prime(r)}{r^2\alpha^\prime(r)-r}}.
\end{align}
The orbital energy per unit mass is
\begin{equation}
\label{energyangular}
\hat{E}=\frac{E}{m_D}=-g_{tt}\, u^t=\sqrt{\frac{e^{2\alpha(r)}\left[1-re^{2\beta(r)}f^r_{\rm cons}\right]}{1-r \alpha^\prime(r)}}.
\end{equation}
From Equations \eqref{velocity-t} and \eqref{energyangular}, we see that $u^t$ and  $\hat{E}$ are approximately the functions of $r$ only.
To derive the orbital decay for the circular orbit, we need to consider the dissipative part of external forces, so we apply the energy balance condition,
\begin{equation}
\label{eq:drdt}
\begin{split}
      \frac{dE}{dt}
      &=m_D\frac{d\hat{E}}{dr}\frac{dr}{dt}+\dot{m}_D\hat{E}\\
      &=\left\langle \frac{d E}{dt} \right\rangle_\text{GW}+\bm{v}\cdot\bm{F}_{\rm DF}+P_a,
\end{split}
\end{equation}
Combining Equations \eqref{eq:F_DF}, \eqref{eq:mDdot}, \eqref{eq:Pa},  \eqref{eq:F_RR}, and \eqref{eq:drdt},
we can can numerically solve the orbital radius $r(t)$ and the mass $m_D(t)$ for the PBH.
With $r(t)$, we can get the evolution of the orbital frequency,
\begin{equation}
\label{orb-frk}
\Omega=\frac{d\phi}{dt}=\frac{u^{\phi}}{u^t}=\sqrt{\frac{e^{2\alpha(r)}[\alpha^\prime(r)-e^{2\beta(r)f^r_{\rm cons}}]}{r-r^2e^{2\beta(r)}f^r_{\rm cons}}}.
\end{equation}

\begin{figure}
\includegraphics[width=0.99\columnwidth]{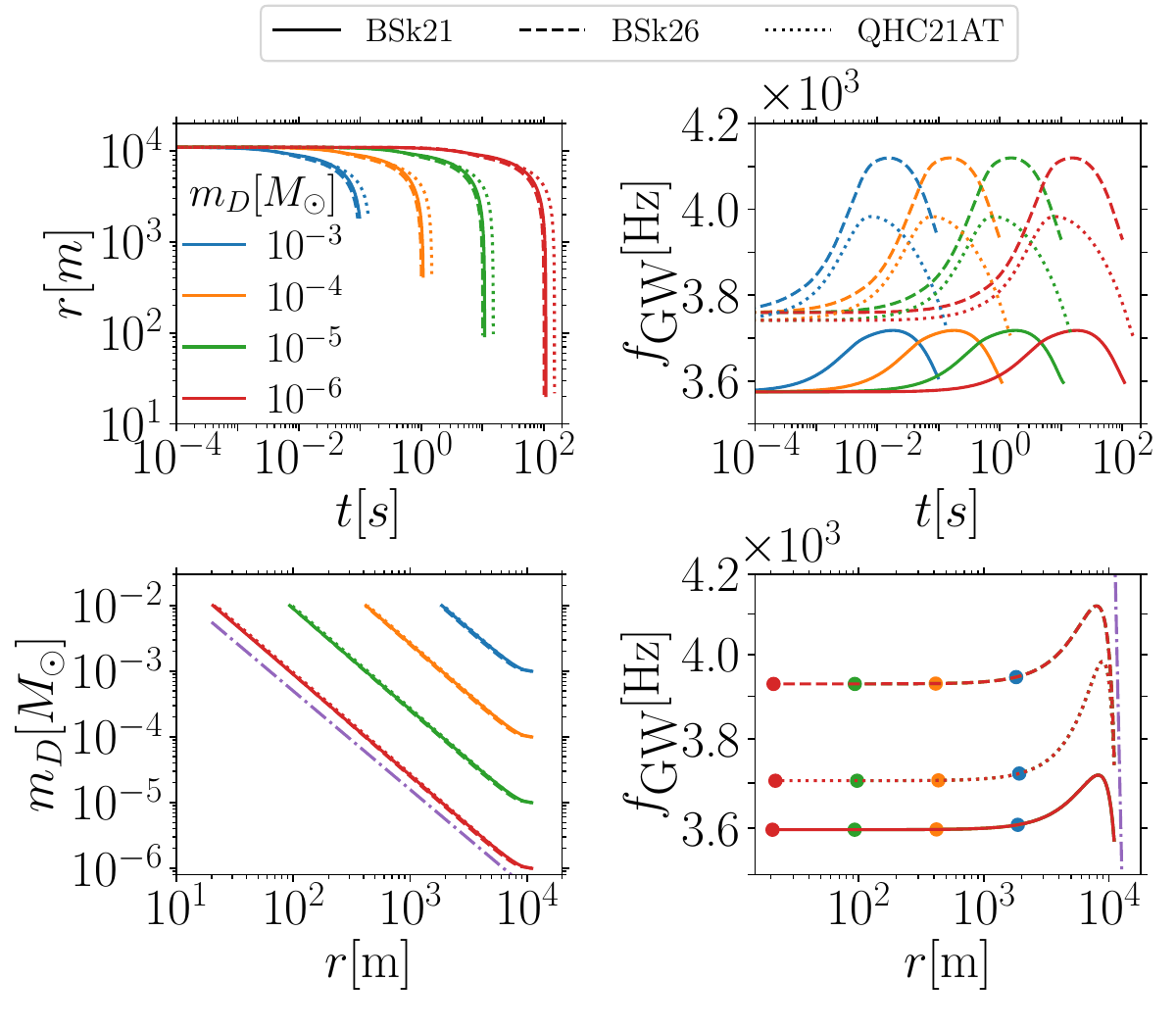}
\caption{
The evolution of the orbital radius $r(t)$, the PBH mass $m_D$, and GW frequency $f_\text{GW}$ for different EOS models.
The initial masses of PBHs are chosen as $10^{-6}$ (red line), $10^{-5}$ (green line), $10^{-4}$ (yellow line), and $10^{-3}\ M_\odot$ (blue line).
The evolution of PBHs begins at $r=11$ km and stops when $m_D=0.01\ M_\odot$.
The purple dash--dotted line in the lower panels is the asymptote $\propto r^{-3/2}$.
In the lower right panel, the dots with different colors denote the GW frequencies at $m_D=0.01~M_\odot$.
}
\label{Fig:OrbitEvolution}
\end{figure}

We numerically calculate the orbital motions of PBHs inside the NS by taking dynamical friction, accretion, and GW radiation into account.
We start the evolution at $r=11$ km ($t=0$) and stop either at the center or when the mass of PBH reaches the value $m_D=0.01~M_{\odot}$ ($t=t_\text{end}$)
as massive PBHs will engulf the NS.
The evolution of GW frequency $f_\text{GW}=\Omega/\pi$,
the orbital radius $r(t)$, and the PBH mass $m_D(t)$ are shown in Figure \ref{Fig:OrbitEvolution}.

As PBHs inspiral toward the NS core, the emitted GW frequency rises to a maximum before decreasing to a constant.
The frequency peaks at $\{8.21, 7.96, 8.90\}$ km for the EOS models BSk21, BSk26, and QHC21AT,  respectively.
Outside the NS core,
the mass function $m(r)$ is almost a constant,
so initially, the GW frequency increases as $r$ decreases.
As the PBH inspirals deeper inside the NS core,
$m(r)$ decreases in proportion to $r^3$.
$\alpha(r)$ and $p(r)$ decrease to a constant as $r$ decreases,
causing the GW frequency to gradually decrease to a constant as $r$ decreases.
Since PBHs follow a geodesic path during each revolution under the adiabatic approximation and the influence of the external force is extremely weak for a single revolution,
the orbital frequency is determined solely by the geodesic motion in the interior metric of NSs,
treating PBHs in the EMRIs as point particles.
Consequently, the GW frequency, which is twice the orbital frequency, is approximately a function of $r$ only and independent of the mass of PBHs, as shown in Figure \ref{Fig:OrbitEvolution}.

Because we have the metric solution within the NS,
we can utilize the behavior of frequency to comprehend the internal structure of NSs and the nuclear matter in their crust and core.
Thus, the unique characteristics of GW frequency act as the smoking gun for GW signals emitted by PBHs inspiraling inside NSs,
and these GW signals can be employed to investigate the internal structure of NSs and the nuclear matter in their crust and core.

From Figure \ref{Fig:OrbitEvolution}, we see that the evolution timescale is inversely proportional to the PBH mass,
with the evolution time becoming shorter as the density of NS increases
As shown in Eqsuations \eqref{eq:F_DF}, \eqref{eq:mDdot}, and \eqref{eq:F_RR},
the leading terms of the dynamical friction, accretion, and GW reaction are all proportional to $m_D^2$,
resulting in a shorter evolution timescale for larger PBH mass.
The leading terms of both dynamical friction and accretion are proportional to $\rho$,
thus the evolution timescale is shorter for the EOS model with higher density.
For PBHs of the same mass, the orbit of the PBH inside BSk26 NS decays the fastest, followed by the orbit inside BSk21 NS,
while the orbit inside QHC21AT NS decays the slowest.
Variations in PBH masses and EOS models result in different orbital evolutions,
which will be reflected in the GWs emitted by the system.

\section{Gravitational waves}
The GW in the transverse-traceless (TT) gauge is
\begin{equation}
h_{i j}^{\mathrm{TT}}=\frac{2}{d_L}\left(P_{i l} P_{j m}-\frac{1}{2} P_{i j} P_{l m}\right) \ddot{\mathcal{I}}^{l m},
\end{equation}
where $d_L$ is the luminosity distance to the source,
and $P_{ij}=\delta_{ij}-n_i n_j$ is the projection operator acting onto GWs with the unit propagating direction $n_j$.
The GW strain measured by the detector is
\begin{equation}\label{signal}
h(t)=h_{+}(t) F^{+}(t)+h_{\times}(t) F^{\times}(t),
\end{equation}
where $F^{+,\times}(t)$ are the interferometer pattern functions, and $h_{+,\times}$ are the two GW polarizations,
\begin{equation}
\label{hamp}
\begin{split}
	h_+(t)=&\frac{4r(t)^2}{d_L}\frac{1+\cos^2\iota}{2}\bigg[\Omega(t)^2 \cos\left(2\varphi(t)\right)m_D\\
 &+\Omega(t) \sin\left(2\varphi(t)\right) \dot{m}_D \bigg],\\
	h_\times(t)=&-\frac{4r(t)^2}{d_L}\cos\iota\bigg[\Omega(t)^2\sin\left(2\varphi(t)\right)m_D\\
 &-\Omega(t) \cos\left(2\varphi(t)\right) \dot{m}_D \bigg],\\
\end{split}
\end{equation}
$\iota$ is the inclination angle between the binary orbital angular momentum and the line of sight, and the orbital phase $\varphi(t)$ is
\begin{equation}
    \varphi(t)=\varphi_0+\int_0^t\Omega(t)dt,
\end{equation}
$\varphi_0$ is the initial orbital phase at $t=0$.

\begin{figure}
\includegraphics[width=0.7\columnwidth]{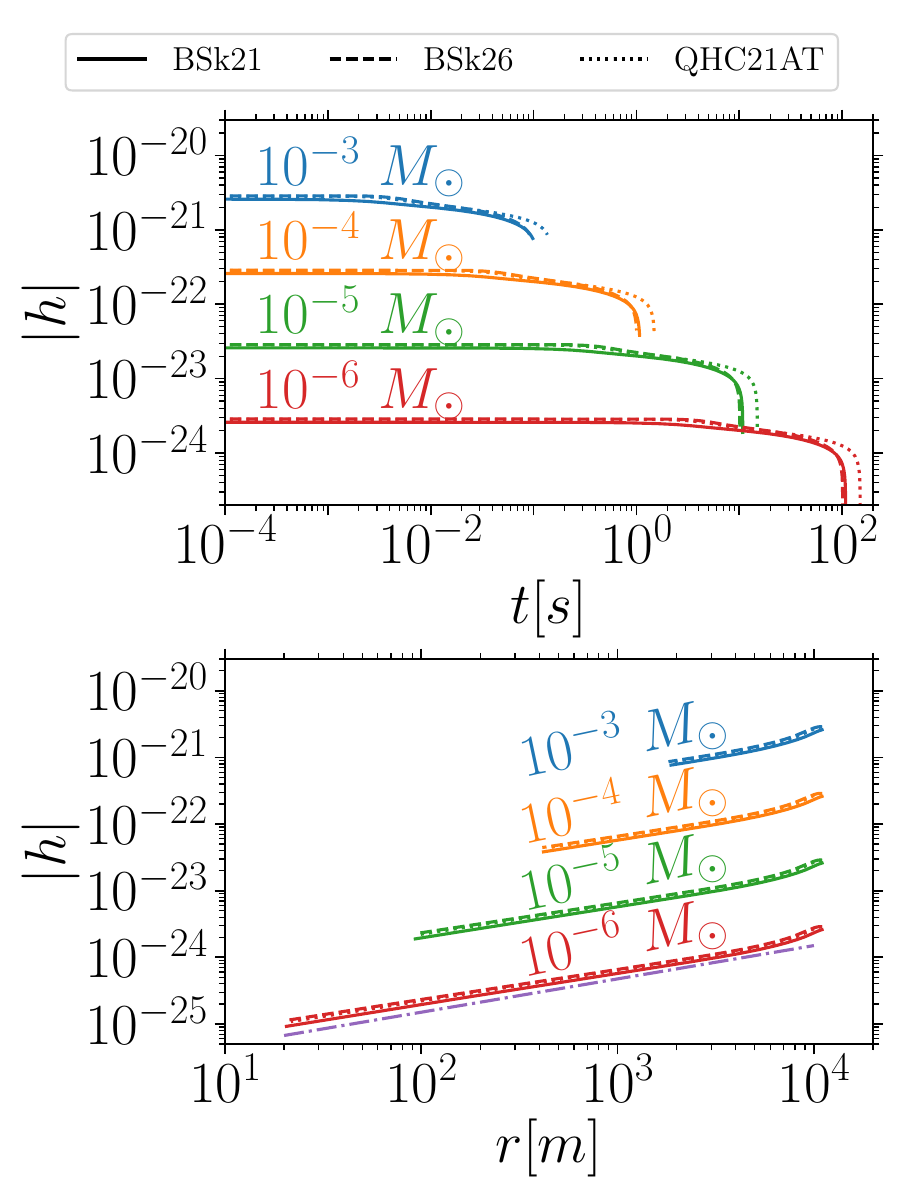}
\caption{
Amplitudes of GWs from PBHs inside NSs for different EOS models.
The initial masses of PBHs are chosen as $10^{-6}$ (red line), $10^{-5}$ (green line), $10^{-4}$ (yellow line), and $10^{-3}\ M_\odot$ (blue line).
The GW strain is $h=h_+-ih_\times$, the inclination angle $\iota=\pi/3$, and $d_L=10\ \text{kpc}$.
The evolution of PBHs starts at $r=11$ km and stops when $m_D=0.01\ M_\odot$.
The purple dash--dotted line in the lower panel is the asymptote $|h|\propto \sqrt{r}$.
}
\label{Fig:GW_Amp}
\end{figure}

The amplitudes of GWs from PBHs inspiraling inside NSs are shown in Figure \ref{Fig:GW_Amp}.
From Figure \ref{Fig:GW_Amp}, we see that the amplitudes of GWs are proportional to the PBH mass and are larger for the EOS model with a higher density.
As seen from Equation \eqref{hamp}, the amplitudes of GWs are proportional to $m_D(\Omega r)^2\sim m_D m(r)/r$.
Since $m_D$ and $m(r)$ are almost constant outside the core,
the amplitudes of GWs decrease as $r^{-1}$.
In the NS core,
since $m_D$ decreases as $r^{-3/2}$ as shown in Figure \ref{Fig:OrbitEvolution}, and $m(r)$ increases as $r^3$,
the amplitudes of GWs increase as $\sqrt{r}$.

\begin{figure}
\includegraphics[width=0.8\columnwidth]{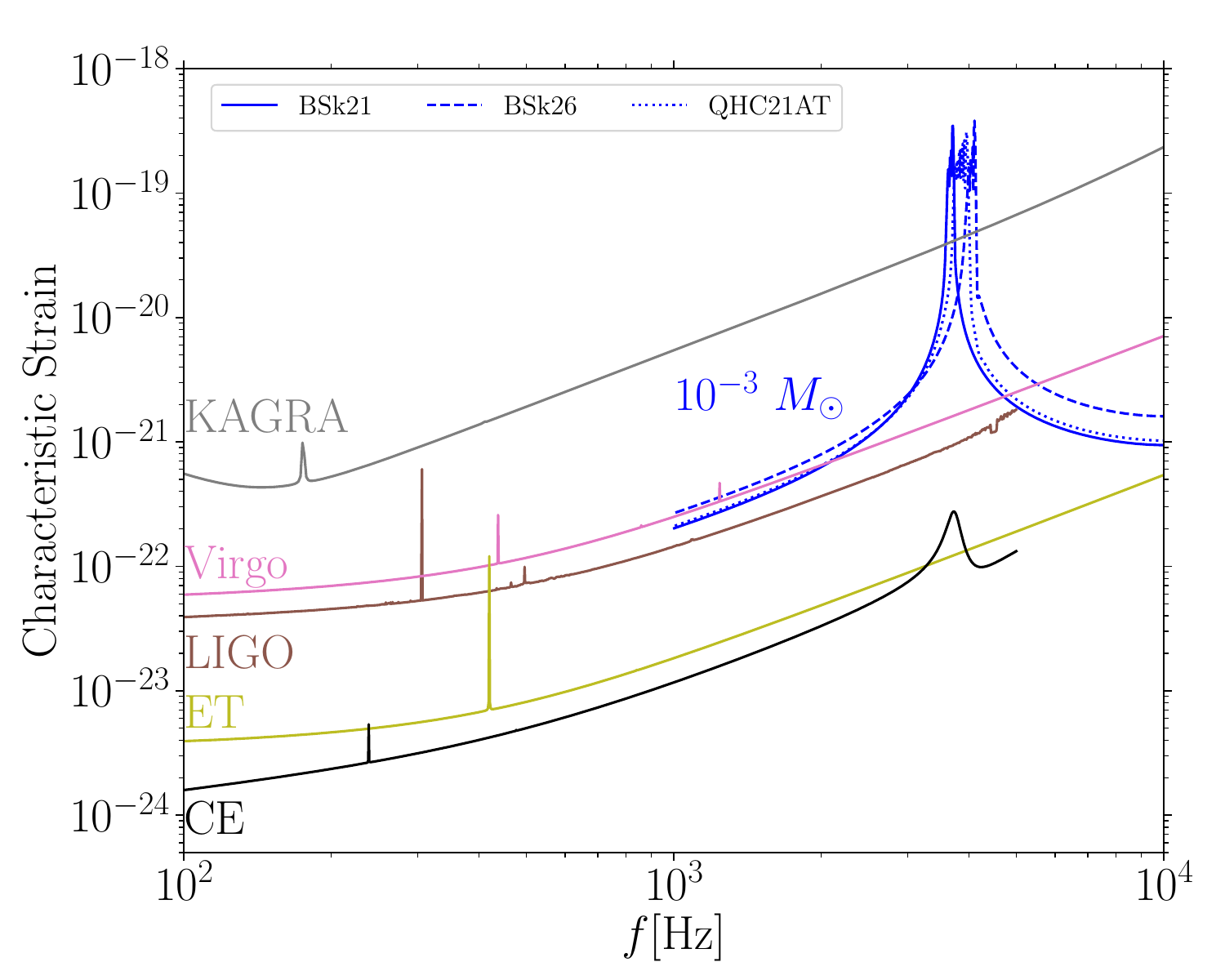}
\caption{
The noise curves along with various sources.
The solid gray, pink, brown, yellow, and black lines denote the noise curves for KAGRA, Virgo, LIGO, ET, and CE, respectively.
The blue lines are for the characteristic strains of a PBH inspiraling inside an NS.
The initial mass of the PBH is $10^{-3}\ M_\odot$.
The solid, dashed and dotted blue lines represent the EOS models BSk21, BSk26, and AHC21AT, respectively.
}
\label{Fig:asd}
\end{figure}

To quantify the difference between two GW waveforms, we use the faithfulness between the two signals,
\begin{equation}\label{eq:def_F}
\mathcal{F}_n[h_1,h_2]=\max_{\{t_0,\phi_0\}}\frac{\langle h_1\vert
	h_2\rangle}{\sqrt{\langle h_1\vert h_1\rangle\langle h_2\vert h_2\rangle}}\ ,
\end{equation}
where $(t_0,\phi_0)$ are time and phase offsets \citep{Lindblom:2008cm}.
Based on the Parseval's theorem,
the noise-weighted inner product between two templates $h_1$ and $h_2$ is defined as
\begin{equation}\label{product}
\left\langle h_{1} | h_{2}\right\rangle=\int_{0}^{t_{\rm end}} \frac{h_1(t) h_2(t)}{S_n(t)} dt=4 \Re \int_{f_{\min }}^{f_{\max }} \frac{\tilde{h}_{1}(f) \tilde{h}_{2}^{*}(f)}{S_{n}(f)} df,
\end{equation}
where $\Re$ means the real part, $\tilde{h}_{1}(f)$ is the Fourier transform of the time-domain signal $h_1(t)$,
$\tilde{h}_{1}^{*}(f)$ is the complex conjugate,
$f_\text{min}$ and $f_\text{max}$ correspond to the minimum and maximum frequencies during the inspiral,
and $S_n(f)$ is the noise spectral density for the GW detector.
Here we chose the noise spectral density (ET-D) of ET as an example \citep{Hild:2010id}.
The noise curves for ET \citep{Hild:2010id} along with CE \citep{Reitze:2019iox}, advanced LIGO \citep{LIGOScientific:2014pky},
advanced Virgo \citep{VIRGO:2014yos}, and KAGRA \citep{KAGRA:2020agh} in the fourth observing run\footnote{https://dcc.ligo.org/T2200043-v3/public},
as well as the characteristic strains for sources with an initial PBH mass of $10^{-3}\ M_\odot$,
are shown in Figure \ref{Fig:asd}.
The signal-to-noise ratio (SNR) is given by
\begin{equation}
\label{eq:snr}
\text{SNR}=\left\langle h|h \right\rangle^{1/2}.
\end{equation}

Table \ref{Tab:faithfulness} shows the faithfulness between GWs using different EOS models.
The faithfulness is larger for the PBH with a larger mass and shorter evolution timescale.
Two signals are distinguishable by the detector if the faithfulness $\mathcal{F}_n<1-N_p/\rm{SNR}^2$, where $N_p$ is the number of model parameters \citep{Flanagan:1997kp,Lindblom:2008cm,McWilliams:2010eq,Chatziioannou:2017tdw}.
If SNR=10, then $1-N_p/\rm{SNR}^2\sim 0.15$.
From Table \ref{Tab:faithfulness}, we see
that $\mathcal{F}_n$ in all the cases we considered is smaller than $1-N_p/\rm{SNR}^2$,
allowing us to use GWs from PBHs inspiraling inside NSs in the Galaxy detected by ET, if they exist, to distinguish EOS models BSk21, BSk26, and QHC21AT.
For PBHs with the initial mass $10^{-6}~M_\odot$ at $d_L=10$ kpc,
the SNRs are too low for detection. Therefore, we reduce the luminosity distance to $3$ kpc to ensure detectability.

\begin{table}
\begin{center}	
\caption{
The faithfulness between GWs from PBHs with different initial masses inspiraling inside NSs with different EOS models detected by ET}
\label{Tab:faithfulness}
\begin{tabular}{|c|c|c|c|c|c|}
\hline
		$m_D$ & BSk21-& BSk21- & BSk26- & SNR   \\	
  $[M_\odot]$ & BSk26 & QHC21AT &QHC21AT &    \\	
        \hline
		$10^{-3}$  & 0.02601 & 0.05220 & 0.16759 & 85/94/93   \\
		\hline
		$10^{-4}$ & 0.00549 & 0.00706 & 0.15112 & 27/28/30   \\
		\hline
		$10^{-5}$ & 0.00075 & 0.00128 & 0.08392 & 9/9/10  \\
		\hline
		$10^{-6}$ & 0.00008 & 0.00010 & 0.02494 & 10/10/10 \\
		\hline
\end{tabular}
\end{center}

\vspace{5pt}
{\bf Note.}
The inclination angle $\iota=\pi/3$, and $d_L=10\ \text{kpc}$ for the initial masses $10^{-3}$, $10^{-4}$ and $10^{-5}$ $M_\odot$.
For PBHs with the initial mass $10^{-6}~M_\odot$, the luminosity distance $d_L$ is set to be $3$ kpc to ensure the detectability with ET.
The evolution of PBHs starts at $r=11$ km and stops when $m_D=0.01\ M_\odot$.
The three values of SNR in the last column are for the results with the three EOS models BSk21, BSk26, and QHC21AT, respectively.

\end{table}

\section{Conclusions}
The concept of using GWs emitted from planet-mass PBHs inspiraling inside NSs to directly probe the interior of NSs for insights into NS EOSs, NS population, internal NS structure, nuclear physics, and the fundamental principles of QCD is quite intriguing.
A robust proof of concept needs to be developed.
We numerically solve the Einstein equations using the EOS of NSs,
and we find that
the transitions from the inner crust to the core occur at $\{11.5,10.8,10.9\}$ km for the EOS models BSk21, BSk26, and QHC21AT,  respectively.
By providing the interior metric of NSs and incorporating the effects of dynamical friction, accretion, and quadrupole-octupole GW reactions,
we generate the most accurate GW templates for  PBHs inspiraling inside NSs.
As PBHs inspiral toward the NS core,
unlike the nearly constant GW frequency anticipated from Newtonian gravity,
the GW frequency initially increases to a maximum before decreasing to a constant value.
The frequency peaks at $\{8.21, 7.96, 8.90\}$ km for the EOS models BSk21, BSk26, and QHC21AT,  respectively.
Furthermore, the GW frequency adheres to the same $f_\text{GW}(r)$ trajectory for PBHs with different masses because PBHs follow a geodesic path during each revolution under the adiabatic approximation,
and the influence of external forces is extremely weak for a single revolution.
The behaviors of the time dependence and distance dependence of the GW frequency differ significantly from those in a binary system during the inspiral phase.
With the metric solution established within the NS, we can leverage the frequency behavior to gain insights into the internal NS structure and the nuclear matter present in the crust and core of NSs.

Near the NS surface, the amplitudes of GWs decrease as $r^{-1}$.
Deep inside the NS, the amplitudes of GWs increase as $\sqrt{r}$ due to the accretion of PBHs.
We also use the faithfulness between GWs from PBHs with masses $\left\{10^{-3},10^{-4},10^{-5},10^{-6}\right\}\ M_\odot$ captured by NSs, as described by EOS models BSk21, BSk26, and QHC21AT, to distinguish different EOS models.
GWs generated from the hypothetical event where a PBH is captured by an NS and falls into it can distinguish different EOS models, such as BSk21, BSk26, and QHC21AT.

We conclude that ET can use GWs from PBHs inspiraling inside NSs in the Galaxy, if they exist, to distinguish EOS models BSk21, BSk26, and QHC21AT.
In other words, measuring the EOS of NSs will be feasible with next-generation GW detectors, and the unique characteristics of GW frequency can be employed to probe the internal NS structure, NS population, and physics under extreme conditions.
The event rates for such hypothetical and speculative scenarios are expected to be low, so detection with current GW detectors is unlikely \citep{Chen:2024qke}.

\section*{Acknowledgments}

This research is supported in part by the National Key Research and Development Program of China under grant No. 2020YFC2201504,
the National Natural Science Foundation of China under grant No. 12175184,
the Chongqing Natural Science Foundation
under grant No. CSTB2022NSCQ-MSX1324, and the China Postdoctoral Science Foundation under grant No. BX20220313 and the China Postdoctoral Science Foundation under grant No. 2023M742297.






\end{document}